\documentclass[preprint,pra]{revtex4}
\usepackage{graphicx}
\usepackage{latexsym}

\def\<{\langle}
\def\>{\rangle}
\def\tr{{\rm tr}}

\def\half{{\frac{1}{2}}}
\def\be{\begin{equation}}
\def\ee{\end{equation}}
\def\bea{\begin{eqnarray}}
\def\eea{\end{eqnarray}}

\def\barA{{\bar{A}}}

\begin{document}
\preprint{cond-mat} \title{Modular Hamiltonian and Typicality of Degenerate States}

\author{Gregory C. Levine and Gon{\c{c}}alo C. Oliveira}
\address{Department of Physics and Astronomy, Hofstra University, Hempstead, NY 11549} 

\date{\today}


\begin{abstract}
Free fermions exhibit Eigenstate Typicality (ET)  meaning that the reduced density matrix, $\rho$, of a small spatial subsystem of a {\sl single}, typical, excited energy eigenstate is gaussian with a modular hamiltonian, $h \equiv -\log \rho$, asymptotically proportional to the free fermion hamiltonian. In effect, the ET may be thought of as providing an independent route to establishing the Fermi-Dirac (FD) distribution for free fermions.  We revisit this argument  for fermions belonging to a superposition of degenerate eigenstates.  Specifically, we show that the modular hamiltonian corresponding to a superposition of degenerate eigenstates in a 1-d free fermion lattice system is interacting, and therefore the reduced density matrix for the subsystem is not gaussian. Measurements of the single particle distribution of a small subsystem would then exhibit non-thermal deviations from the FD distribution.  These features may be traced to a violation of Wick's theorem for the chosen excited state.  We develop a perturbation theory for the modular hamiltonian corresponding to a weak mixture of degenerate states and study the modular hamiltonian (exactly) computationally. Finally, we speculate that initial conditions in a realistic fermi system evolving towards equilibrium might be encoded in the interaction terms of the modular hamiltonian and consequent correlations of fermions.
\end{abstract}

\maketitle
\section{Introduction}

The Eigenstate Thermalization Hypothesis (ETH) \cite{1991PhRvA..43.2046D,1994PhRvE..50..888S,2018RPPh...81h2001D,2016AdPhy..65..239D,2008Natur.452..854R} asserts that if a measurement of a local quantum operator is made in a small spatial subregion of a single, typical, excited energy eigenstate of a many body hamiltonian, the results will match that of a comparable measurement made within the conventional thermal ensemble at a temperature comparable to the energy of the eigenstate. Considering the time evolution of a system as it equilibrates, if some initial condition is realized by mixing a set of neighboring energy eigenstates together, the time evolution acts to select a small number of adjacent eigenstates in the long time limit.  If all eigenstates belonging to the set behave identically, then the expectation value of a local operator will be independent of the specific initial conditions.  This resolves the historic tension between equilibration in classical mechanics---which relies upon chaos and mixing to bring about independence of observables from initial conditions---and quantum evolution of a finite system which does not have a comparable mechanism to lose information about initial conditions.

 Closely related is the statement of Canonical Typicality (CT)---that the reduced density matrix of a small subsystem belonging to a pure state $|\Psi\>$ reproduces thermal expectation values at a temperature related to the energy density of $|\Psi\>$ \cite{2006NatPh...2..754P,2006PhRvL..96e0403G, 2018PhRvX...8b1026G, 2015PhRvB..91h1110L}.  The work in this manuscript concerns the description of the subsystem in cases in which the pure state spans a large number of degenerate eigenstates. 

"Typicality" and the ETH continues to attract much attention for several reasons---its primary role in deepening our understanding of equilibrium in thermodynamics (see, for example, \cite{2024PhRvX..14d1051M, 2023PhRvA.107c2215L, 2018PhRvE..98f0103T,2022PhRvB.105c5106L}), but also somewhat unexpectedly in the realm of quantum information and quantum error correction codes \cite{2019PhRvL.123k0502B,2019JHEP...08..152B}. Both modular flow---the effective time evolution imposed by the modular hamiltonian---and quantum error correction codes, in turn, are central to the current understanding of bulk/boundary operator reconstruction in the AdS/CFT \cite{2015JHEP...04..163A,2016JHEP...06..004J, 2016PhRvL.117b1601D, 2017JHEP...07..151F}. 

To frame the topic of this manuscript, consider a one-dimensional many body system partitioned into two regions, $A$ and its complement $\bar{A}$.  If a quantum state of the system $|\psi \>$ is constructed, the modular hamiltonian in region $A$ corresponding to this state is defined to be 
\be
h \equiv -\log \rho {\,\,\,\rm where\,\,\,} \rho \equiv \tr_\barA |\psi \>\<\psi|
\label{equation_0}
\ee
 Since the reduced density matrix, $\rho = \exp{(-h)}$ formally resembles the Boltzmann probability distribution of equilibrium statistical mechanics, $\exp{}(-\beta H)/\tr \exp{(-\beta H)}$ ($\beta$ is the inverse temperature and $H$ the hamiltonian of the system), it is sometimes possible to represent the modular hamiltonian as a local hamiltonian divided by a local {\sl effective} temperature even if the physical temperature is zero.  There are only a few cases where $h$ may be computed analytically exactly. If $|\psi\>$ is the ground state of a noninteracting system and $A, \barA$ represent an equal (infinite) bi-partition of the system, $h$ is the Rindler hamiltonian---roughly, the original local, noninteracting, hamiltonian divided by an effective temperature that diverges at the boundary between $A, \barA$. As this is a replica calculation, the modular hamiltonian for related, finite, subsystem geometries may be obtained as well by conformal mapping if the system is a CFT \cite{1994NuPhB.424..443H, 2011JHEP...05..036C}. Another example is afforded by the ET itself. The reduced density matrix of a small subregion of a typical many-body excited state, having energy $e$, will have a modular hamiltonian given approximately \cite{2018PhRvE..97a2140D, 2018JSMTE..03.3101L} by $\beta_e H$ where $H$ is the hamiltonian of the original system and $\beta_e$, an inverse temperature derived from the energy of the state, $e$ \cite{2023PhRvE.107b4102B}.  The modular hamiltonian of disconnected subregions in a CFT was explored in \cite{2009CQGra..26r5005C}. 
Finally, a general framework for computing the modular hamiltonian and relative entropies in CFTs for excited states corresponding to primary operators has been developed in \cite{2016PhRvL.117d1601L, 2020PhRvD.102b6021A}. 

Apart from the role of the modular hamiltonian in statistical mechanics, the modular hamiltonian and modular {\sl flow}---the effective, Heisenberg-like, time evolution accorded a modular hamiltonian---have been recently shown to play an essential role in the AdS/CFT and its interpretation in terms of quantum error-correction codes \cite{2015JHEP...04..163A,2016JHEP...06..004J, 2016PhRvL.117b1601D, 2017JHEP...07..151F}. These and other related works establish the notion of subregion-subregion duality (reconstruction of bulk operators in a Cauchy slice from CFT operators in that slice), the exact form of Rindler reconstruction of bulk operators from the modular flow of boundary operators and quantum-correction as a feature of the bulk, bringing a new consistency to the AdS/CFT. Briefly, these works rely on a type of equivalence between the modular hamiltonian of a CFT subregion and the modular hamiltonian projected onto the "code" subspace representing the bulk degrees of freedom. 

Because of the role of the modular hamiltonian in these several active areas, we believe it is important to study its construction beyond the known cases. The case we focus upon is that of the modular hamiltonian corresponding to the superposition of degenerate free fermion states where non-locality and interactions in $h$ appear to be unavoidable.  In particular, an excited states of a fermi system relevant to the ETH  belongs to a large---possibly extensive---set of degenerate states. The difficulty in the resulting construction naturally follows from the definition of the modular hamiltonian as a logarithm of an operator; consider the following simple case: A state $|\psi\>$ is defined to be a superposition of two normalized, excited energy eigenstates:
\be
|\psi\> = \cos{\frac{\theta}{2}}|e_a\> + \sin{\frac{\theta}{2}} |e_b\>
\ee
As a consequence of Wick's theorem, the modular hamiltonian, $h_a$, corresponding to any {\sl pure} eigenstate $|e_a\>$ of a non-interacting hamiltonian is non-interacting (quadratic) itself \cite{2003JPhA...36L.205P}.
\be
h_a = -\log \rho_a \equiv -\log \tr_\barA |e_a\>\< e_a|
\ee

Expressing corrections to $h_a$ perturbatively in $\theta$ and ignoring the complications arising from the Baker-Campbell-Hausdorff (BCH) expansion of the logarithm of matrix products, one then finds
\be
\delta h_a \approx -\theta \rho_a^{-1}\tr_\barA(|e_a\>\<e_b| + |e_b\>\<e_a| )
\ee
If $|e_a\>$ and $|e_b\>$ are related to one another by transitions in the momenta of the constituent particles, we will show that this relationship between the two excited states gives rise to interactions and, quite possibly, non-locality in $\delta h_a$ that are not present in $h_a$ or $h_b$.  The choice of degenerate eigenstates of course guarantees that the resulting reduced density matrix is time-independent and potentially represents a stationary equilibrium state.  Furthermore, if the $\barA$ region is vastly larger than the $A$ region, mixing typical degenerate states appears to lead to a modular hamiltonian with effectively random interactions. 
 
 In this manuscript we study the modular hamiltonian for excited states of noninteracting spinless fermions, both computationally (exactly) and analytically by perturbation theory. When superpositions of pure states are formed, Wick's theorem cannot be applied in calculating $\rho$ and $h$ even though the fermions are noninteracting, and the numerical calculations must be performed in the full many-body Hilbert space (with careful treatment of the singular nature of matrix logarithms).  In section 2, we review the argument for free fermions satisfying the ET and the computation of the modular hamiltonian for superpositions of degenerate states demonstrating the presence of interactions.  In section 3, we develop a perturbation theory to understand the origin of these features. In section 4, we discuss the physical applicability and consequences for realistic fermion systems. In particular, the appearance of interactions must be squared with the traditional renormalization group picture of fermions with asymptotically weak low energy interactions.

\section{Eigenstate Typicality for Free Fermions}

Unlike the classical picture of equilibration, the ETH suggests that interactions do not {\sl uniquely} play a role in achieving an equilibrium state. Remarkably, it was shown by Lai and Yang in \cite{2015PhRvB..91h1110L} that a small subsystem within a non-interacting fermion system in a pure state has a thermal description, in effect, due to entanglement and identical particle statistics. They term this effect Eigenstate Typicality (ET). In this section we begin by presenting their argument for ET and follow with how the resulting calculation of the modular hamiltonian changes when the system is in a superposition of pure states, Wicks's theorem is violated and the density matrix no longer assumes a gaussian form.  

Consider the a one-dimensional system of non-interacting, spinless fermions on an $L$-site lattice:
\be
H = -\sum_{j=1}^L{(c^\dagger_j c_{j+1} + c^\dagger_{j+1}c_j ) }
\label{spinless_fermions}
\ee
A pure excited $N$-particle state of $H$ having total energy $e_a$ for may be written:
\be
|e_a\> = \prod_{m=0}^{N-1}{c^\dagger_{k^a_m} |0\>}
\label{pure_state}
\ee
where $c_k$ are the corresponding momentum operators and the particular {\sl ordered} set of occupied momentum states $\{k^a_m\}$ defining $|e_a\>$ are $N$ momentum values drawn from the complete set of possible momenta $k_n = 2\pi n / L$, $n = 0\ldots L-1$. Expectation values of products of operators calculated with any $|e_a\>$ satisfy Wick's theorem and factor into products of one particle correlation functions because a unique normal ordering of operators may be specified with respect to the ordering chosen in (\ref{pure_state}). The factoring of correlation functions allowed by Wick's theorem then leads to the observation \cite{2003JPhA...36L.205P} that the spatial correlation function within $A$ computed for any pure eigenstate $|e_a\>$ must satisfy
\be
m^a_{jl} \equiv \< e_a| c^\dagger_j c_l | e_a\> = \tr_\barA \rho_a c^\dagger_j c_l
\ee
where $\rho_a$ has the gaussian (non-interacting) form:
\be
\rho_a = \exp{(-\sum_{i,j \in A}{h^a_{ij} c^\dagger_i c_j})}
\ee
Represented as matrices within the single particle Hilbert space (boldface is reserved for matrices in the single particle hilbert space), the relationship between the modular hamiltonian for a sub-region $A$ and the correlation function for state $|e_a\>$ follows:
\be
\mathbf{h}_a = \log{(\mathbf{m}_a^{-1}-\mathbf{I})}
\ee
The calculation of $h^a_{ij}$ follows directly from the calculation of the correlation function:
\be
m^a_{jl} = \frac{1}{L}\sum_{m=1}^N{e^{i k^a_m (j-l)}} 
\label{corr_func}
\ee
Lai and Yang argue that a typical excited state $|e_a\>$ has the property that the average occupation numbers for the set $\{ k^a_m\}$ change gradually as a function of $m$. Thus, for the purposes of computing $\mathbf{m}_a$ within a small region $A$ of linear size $L_A << L$, the sum (\ref{corr_func}) may be broken into momentum cells containing $g$ consecutive values of momenta wherein the phase factor appearing in (\ref{corr_func}) is roughly constant. Consider cell number $p$---this cell contains the momenta in the set $k \in[\frac{2\pi}{L}p g, \frac{2\pi}{L}(p+1)g ]$. For the phase factor in (\ref{corr_func}) to remain approximately constant as $\Delta x = j-l$ ranges over $L_A$, it is sufficient to choose $g<< L/L_A$ while keeping $L_A << L$. They then identify the particular momentum occupation of the set  $\{ k^a_m\}$ within a cell as a {\sl microstate} and maximize the total number states subject to the constraints of total energy and number of particles.  Lai and Yang's calculation, in effect, parallels the standard calculation of FD statistics with choice of microstates within a cell being identified with degeneracy.  This is a subtle point in that the microstates are not actually {\sl degenerate} but rather {\sl equivalent}---through typicality---in leading to an effectively equivalent correlation function and modular hamiltonian for a small subregion.  


\begin{figure}[htbp]
    \centering
    \includegraphics[width=1.1\linewidth]{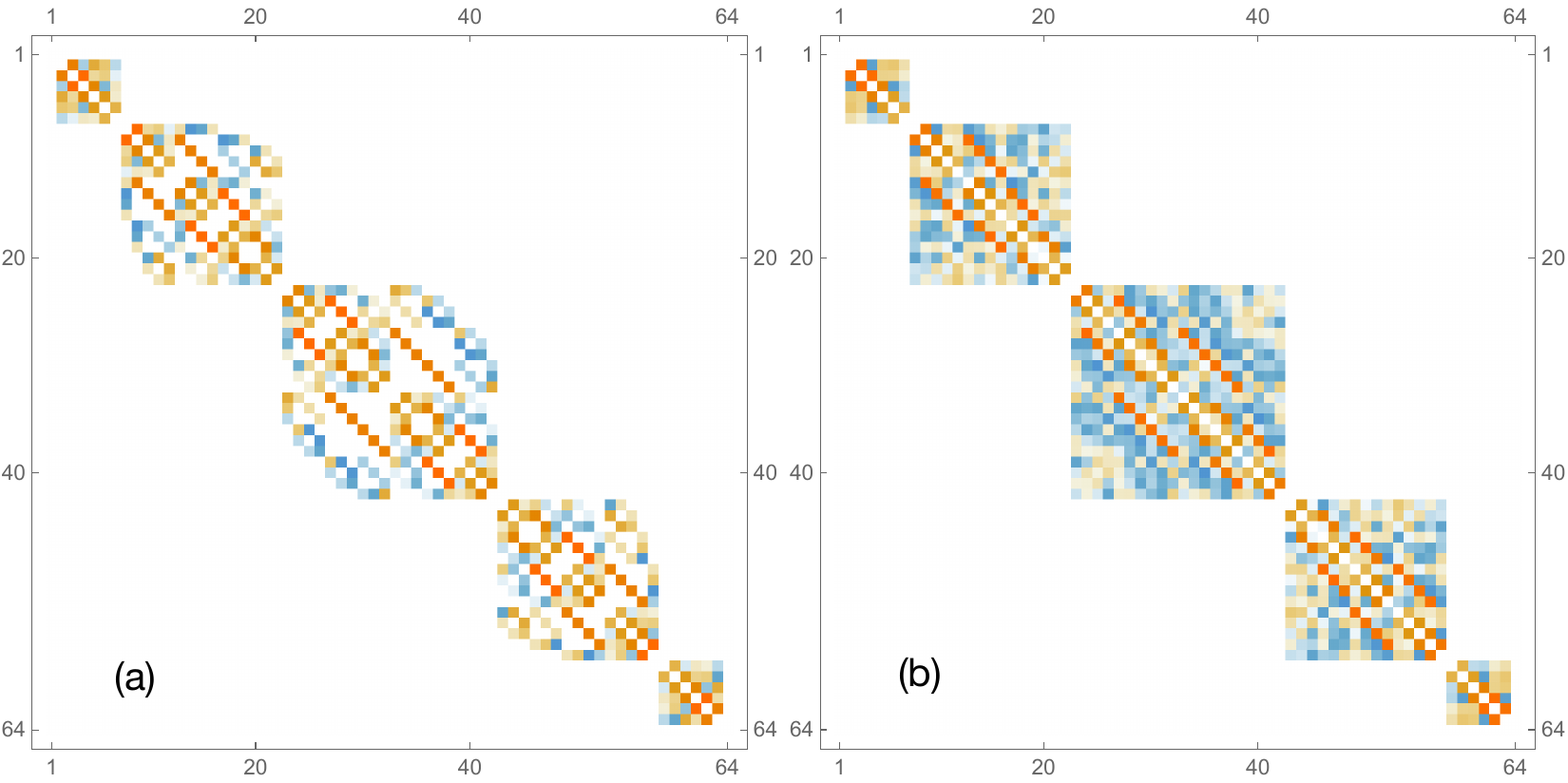}
    \caption{Modular hamiltonian, $h_\theta$, computed in the many-body hilbert space for the state $ |\theta\> = \cos{\frac{\theta}{2}}|e_j\> + \sin{\frac{\theta}{2}} |e_k\> $ where $j=3$ and $k=4$ are the third and fourth eigenstates of the (12/6/6) system sharing energy $e_3=e_4 \approx -6.3459467\ldots$. (a) $h_\theta$ for the pure eigenstate $|\theta = \pi/2\>$ with unique momentum ordering (eq. \ref{pure_state}) satisfying Wick's theorem. The large (orange) off-diagonal terms represent nearest neighbor (nn) and next-nearest neighbor (nnn) (local) transitions demonstrating a free, noninteracting modular hamiltonian. All two-body and higher interaction terms are exactly zero. (b) $h_\theta$ for an "impure" state $|\theta = \pi\>$ lacking unique momentum ordering and thus violating Wick's theorem. Two-body and higher interaction terms are non-zero. The free particle nn and nnn matrix elements persist unchanged when interactions appear in $h_\theta$ as seen in (b). Diagonal elements in both (a) and (b) have been suppressed to enable colored distinction of off-diagonal magnitudes.}
    \label{mod_ham1}
\end{figure}

\begin{figure}[htbp]
    \centering
    \includegraphics[width=1.1\linewidth]{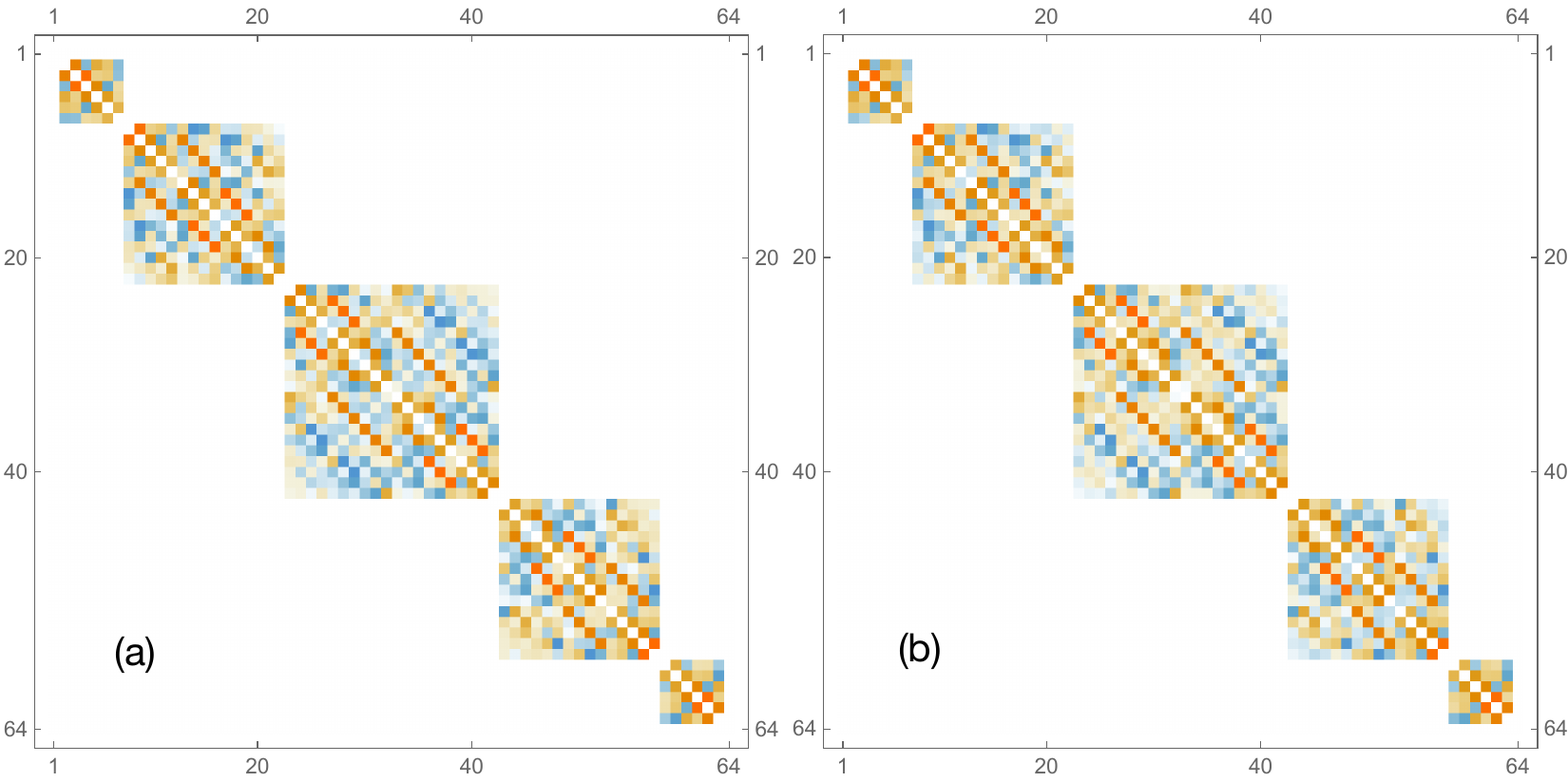}
    \caption{Modular hamiltonian $h_\theta$ for mixing angles close to $\theta = \theta_{\rm pure} = \pi/2$. (a) Mixing angle chosen slightly smaller, $\theta = \frac{\pi}{2} - \epsilon$, and (b) slightly bigger $\theta = \frac{\pi}{2} + \epsilon$ than the pure, unique momentum ordering present in the $\theta = \frac{\pi}{2}$ case ($\epsilon = 0.01\pi$)  Clearly, interacting terms in $h_\theta$ arise under tiny mixture perturbations from the pure case suggesting that an {\sl interacting} $h_\theta$ is generic while a noninteracting $h_\theta$ is highly special.}
    \label{mod_ham2}
\end{figure}
\section{Computations of the Modular Hamiltonian}

It is then natural to ask what the modular hamiltonian actually looks like for superpositions of degenerate states.  To study this question we numerically computed the reduced density matrix and modular hamiltonian for small 1-dimensional systems of lattice fermions described by hamiltonian (\ref{spinless_fermions}). To realize the most "independent" possible subregion, we chose fixed boundary conditions  and a subregion  $A$ of $L_A$ sites with only one boundary. The notation $(L/L_A/N)$ refers to system with $L$-sites, an $L_A$-site subsystem bounded by the fixed boundary and the $L-L_A$-site subsystem $\barA$ and $N$ fermions. We form a superposition state $|\theta\>$ 
\be
|\theta\> = \cos{\frac{\theta}{2}}|e_j\> + \sin{\frac{\theta}{2}} |e_k\>
\label{theta_super}
\ee
where $|e_j\>$ is an excited state degenerate with $|e_k\>$. Since these excited states were obtained by direct diagonalization of the real space hamiltonian (\ref{spinless_fermions}) in the many-particle hilbert space, they are not necessarily individually pure (possessing a unique momentum ordering in the sense of eq. (\ref{pure_state})) although two linear combinations at particular values of $\theta = \theta_{\rm pure}$ must be pure.  We directly compute the reduced density matrix, $\rho_\theta$, and modular hamiltonian, $h_\theta$, for subsystem $A$
\be
  h_\theta =-\log{\rho_\theta} = -\log \tr_\barA|\theta\>\<\theta| 
\label{h_theta_def}
\ee 
For instance in the $(12/6/6)$ system, the chosen state density matrix of dimension $C^{12}_6 = 924$ is traced over $2^6$ bit strings comprising subsystem $\barA$ resulting in a $2^6$ dimension modular hamiltonian for subsystem $A$.  The sensitivity of the matrix logarithm implicit in $h_\theta$ limited us to very small system sizes even when using enhanced precision. In the $(12,6,6)$ system, direct diagonalization of hamiltonian (\ref{spinless_fermions}) keeping a precision of less than 22 digits resulted in small normalization errors in $\tr_A\rho_\theta = 1$ that led to large errors in the off-diagonal matrix elements of $h_\theta$. For periodic boundary conditions, we found that excited eigenstates of (\ref{spinless_fermions}) constructed directly from antisymmetrized analytical single particle wavefunctions  permitted standard double precision (16 digits) computations for the $(12/6/6)$ system. 

Such instabilities are not present in the single particle hilbert space computations of $\rho_a$ from the single particle correlation function and we were able to compare $\theta = \theta_{\rm pure}$ results to the one-body matrix elements of the many-body hilbert space computations of the modular hamiltonian as a cross-check. Another check was simply observing that the two-body (interacting) matrix elements of the modular hamiltonian were identically zero at points $\theta=\theta_{\rm pure}$.

 \begin{figure}[htbp]
    \centering
    \includegraphics[width=1.1\linewidth]{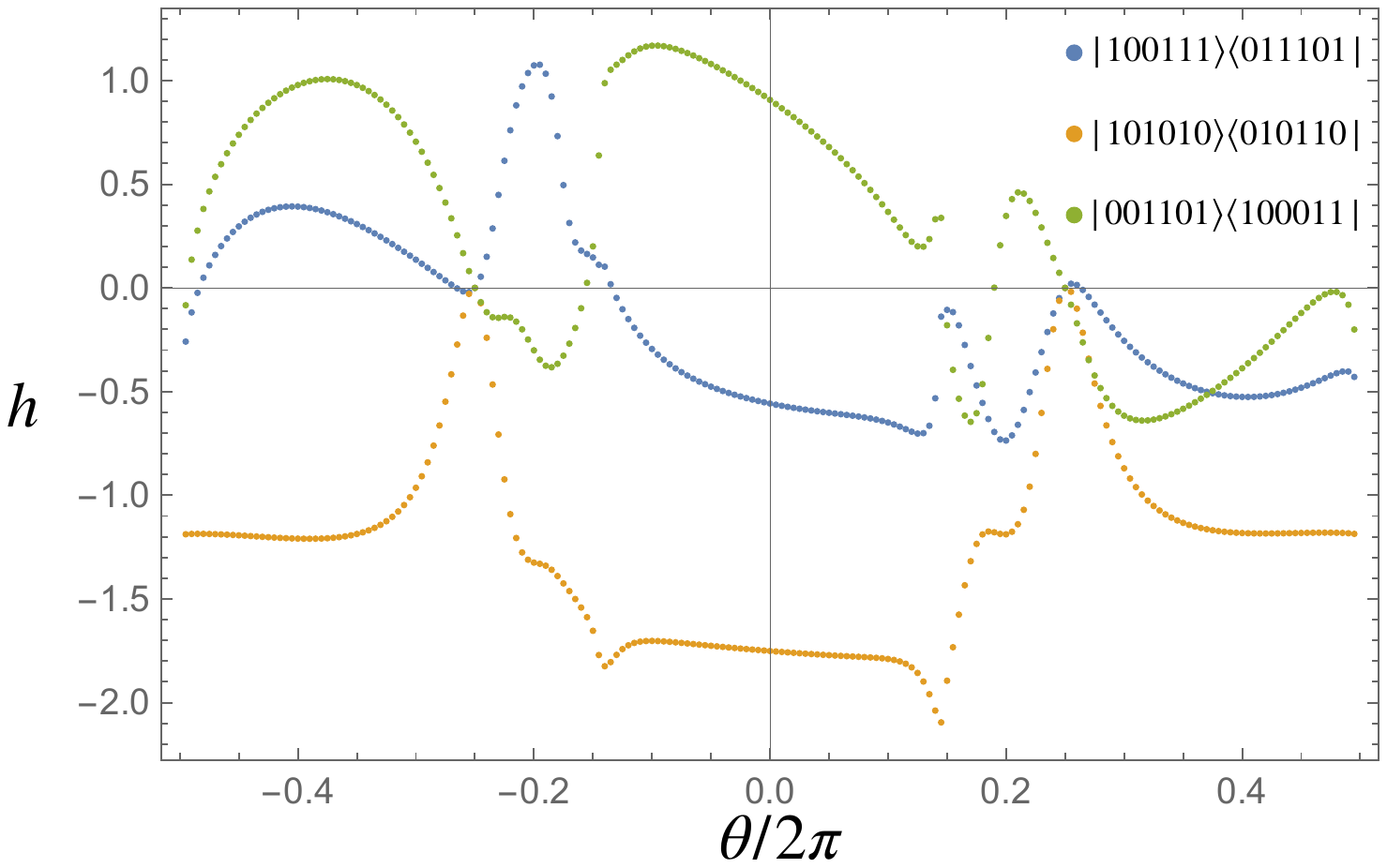}
    \caption{Three representative two-body (interacting) matrix elements of the modular hamiltonian corresponding to the mixture (\ref{theta_super}) of two degenerate states $|e_a\>$ and $ |e_b\>$.  Note that all three matrix elements are zero at $\theta =\theta_{\rm pure} =\pm \pi/2$ where the mixture is pure (in the sense of (eq. \ref{pure_state}) ) with the phases numerically realized in our diagonalizations. Note that one element displayed here passes through $\theta_{\rm pure}$ {\sl linearly} while the other two, quadratically.}
    \label{matrix_elements}
\end{figure}

We will refer to elements $\<f| c^\dagger_j c_l |i\>$ as one-body elements and  $\<f| c^\dagger_j c^\dagger_l c_m c_n |i\>$ as two-body elements where the initial and final states of the element in the $A$ subspace will be presented as bit-strings: (e.g. $|011010 \>$ for the $(12/6/6)$ system). The dimension of the hilbert space for $A$ subsystem on the $(12/6/6)$ case is then $M_A \equiv 2^6$. 

Figure (\ref{mod_ham1}) shows the full modular hamiltonian, $h_\theta$, for a pure state (a) and a mixed state (b) of the $(12/6/6)$ system. Although the specific bit strings are not displayed with the matrix, $h_\theta$ in figure (\ref{mod_ham1}a) had only nonzero one-body terms and thus represents noninteracting fermions. In figure (\ref{mod_ham1}b) the large number of additional nonzero elements correspond to two and four body interactions.  Thus interactions between fermions are effectively induced in a sub-system corresponding to a superposition of pure states. The one-body terms remain largely independent of $\theta$ and do not vanish at the $\theta = \theta_{\rm pure}$ points.  Figure (\ref{mod_ham2}) shows $h_\theta$ for $\theta$ close to $\theta_{\rm pure}$ to illustrate that interactions appear abruptly once momentum ordering (eq. \ref{pure_state}) is not strictly satisfied.  

All $M_A$ initial and final elements were present in the modular hamiltonian consistent with $h_\theta$ obeying particle number conservation within $A$. The two-body matrix elements connecting four sites did not appear to depend on the separation between sites consistent with the anticipated non-locality of the modular hamiltonian (see, for instance, \cite{2017JHEP...07..151F, 2023JHEP...11..097L}) but depended sensitively upon the the mixing angle $\theta$. Even within perturbation theory (described next) it is not clear how non-locality in $O$ directly manifests in non-locality of $h_\theta$.

Figure (\ref{matrix_elements}) shows particular two-body matrix elements of $h_\theta$ for the $(12,6,6)$ system as a function of the mixing angle $\theta$. For  $\theta \ne \pm \pi/2$ two-body elements are always present as demonstrated for this representative set.  The representative matrix elements are seen to have a complicated $\theta$ dependence but always vanish at the $\theta = \theta_{\rm pure}$ points.  Note that one element displayed here passes through $\theta_{\rm pure}$ {\sl linearly} while the other two are quadratic in the vicinity of $\theta_{\rm pure}$.

\section{Perturbation Theory For the Modular Hamiltonian}
Analytic calculations of the modular hamiltonian for pure excited states in CFTs have been achieved in \cite{2016PhRvL.117d1601L} and could in principle be extended to the superpositions described here \cite{nonlocal_prep}. One might expect that off-diagonal two-body terms of $h_\theta$ must correspond in some fashion to real-space two-body scattering reflecting momentum-space transitions $|e_a\>  \leftrightarrow |e_b\>$ appropriately projected onto region $A$. Along these lines, we will show that perturbation theory describes the relationship between real space scattering within the Schmidt basis of $\barA$ and scattering within the Schmidt basis of $A$, the latter associated with interacting terms in the modular hamiltonian.  

For the state (\ref{theta_super}), the reduced density matrix for region $A$ is:
\begin{eqnarray}
\rho_\theta &=& \cos^2{\frac{\theta}{2}} \rho_a+ \sin^2{\frac{\theta}{2}} \rho_b \\ &+& \half \sin{\theta}\, \tr_\barA (|e_a\>\<e_b| + |e_b\>\<e_a|) \\
&=& \rho_a +  \half \theta\, \tr_\barA (|e_a\>\<e_b| + |e_b\>\<e_a|) + {\cal{O}}(\theta^2) 
\label{rdm_theta}
\end{eqnarray}
The results of  \cite{2015PhRvB..91h1110L} imply that the reduced density matrices of all typical excited states are all close to one universal density matrix (the thermal density matrix) and thus $\cos^2{\frac{\theta}{2}} \rho_a+ \sin^2{\frac{\theta}{2}} \rho_b \approx \rho_{a,b}$ for all $\theta$ is a reasonable approximation up to corrections of ${\cal{O}}(\theta^2)$. 

Both linear and quadratic behaviors were observed computationally for two-body (interacting) matrix elements of $h_\theta$  in the vicinity of a pure state (now $\theta_{\rm pure} = 0,\pi$). For perturbation theory we study only the linear terms in $\theta$ for the admixture of degenerate states, and the modular hamiltonian is written:
\be
h_\theta \equiv -\log{\rho_\theta} = -\log{(\rho_a + \frac{\theta}{2} M_{ab})}
\label{h_theta_perturbation}
\ee
where
 \be 
 M_{ab} \equiv \tr_\barA (|e_a\>\<e_b| + |e_b\>\<e_a|)
 \label{M_ab}
 \ee
The logarithm of operators may be expanded in a series of nested commutators by the BCH expansion, but we found this approach complicated. (We note that a very complete perturbation theory has been developed in \cite{2023JHEP...11..097L} for the logarithm of a linear operator; presumably our case does not require this full generality.) Instead, using the integral definition of matrix logarithm,
\be
\log{X} = \int_0^\infty{ [I(1+s)^{-1} -  (X + sI)^{-1}]ds }
\ee
the logarithm of the following matrix combination may be found to $\cal{O}(\epsilon)$ by direct matrix manipulation (for an alternative, see Appendix A)
\begin{eqnarray}
&&\log{(X + \epsilon Y)} = \log{X} \\ &+& \epsilon \int_0^\infty{[(X+ sI)^{-1}Y(X + sI)^{-1}) ]ds} + {\cal{O}}(\epsilon^2)
\label{log_expansion}
\end{eqnarray}
Thus,
\be
h_\theta = h_a - \frac{\theta}{2} \int_0^\infty{[(\rho_a+ sI)^{-1}M_{ab}(\rho_a + sI)^{-1}) ]ds} + {\cal{O}}(\theta^2)
\label{int_h}
\ee

To analyze the expression for $M_{ab}$ (\ref{M_ab}) further, we express the pure states $|e_a\>$ and $|e_b\>$ as related to one another by an operator $O^\dagger$ acting in the full $A,\barA$ hilbert space:
\be
|e_b\> = O^\dagger |e_a\>
\ee
Now the state $|e_a\>$ is expanded in its Schmidt basis,
\be
|e_a\> = \sum_\alpha{d_\alpha |\alpha\> \overline{|\alpha\>}}
\ee
and $M_{ab}$ may be written:
\begin{eqnarray}
M_{ab} &=& \sum_{\alpha^\prime \alpha}{d_\alpha d_{\alpha^\prime}( |\alpha\> \overline{\<\alpha^\prime|} \<\alpha^\prime|O \overline{|\alpha\>} + \overline{\<\alpha^\prime|}O^\dagger |\alpha\> \overline{|\alpha\>} \<\alpha^\prime|  )} \nonumber \\
&=&: \sum_{\alpha^\prime \alpha}{d_\alpha d_{\alpha^\prime}} G_{\alpha \alpha^\prime}
\end{eqnarray}
To arrive at this expression we note that $\tr_\barA $ expresses a sum over a complete set of many body states of the $\barA$ region which is only partially spanned by the set $\{|\bar{\alpha}\> \}$, while the set $\{|\alpha\> \}$ is nominally complete in the $A$ region as a result of eigenstate typicality. However, a complete set $\{|\bar{\beta}\> \}$ might be defined as the Graham-Schmidt completion of the set $\{|\bar{\alpha}\> \}$.  Then the trace $\tr_\barA = \sum \<\bar{\beta}| \ldots  |\bar{\beta} \>$ reduces to the trace $ \sum \<\bar{\alpha}| \ldots  |\bar{\alpha} \>$ when applied to $M_{ab}$ written in the Schmidt basis. Further,  it should be recognized that the operator $O$ which induces transitions between degenerate states is supported in $A,\barA$, though $M_{ab}$ is simply a matrix in the region $A$ hilbert space. For instance, $O^\dagger$ creates an excitation in the full hilbert space Schmidt element $O^\dagger |\alpha\>\overline{|\alpha\>} $ which is projected onto the $\barA$ state $\overline{\<\alpha^\prime|} $ to determine the matrix element $G_{\alpha \alpha^\prime}$. Since $|\alpha\>$ is an eigenvector of both $\rho_a$ and $h_a$,
\begin{eqnarray}
\rho_a |\alpha\> &=& d_\alpha^2 |\alpha\> \\
h_a |\alpha\> &=& E_\alpha |\alpha\>
\end{eqnarray}
where $E_\alpha \equiv - 2\log{d_\alpha}$. Now, the integral in (\ref{int_h}) is expressed in the Schmidt basis for region $A$:
\begin{eqnarray}
 &&\int_0^\infty{[(\rho_a+ sI)^{-1}M_{ab}(\rho_a + sI)^{-1}) ]ds} \nonumber \\ &=&  \int_0^\infty{ds\sum_{\alpha \alpha^\prime}{\frac{d_\alpha}{d_\alpha^2 + s} \frac{d_{\alpha^\prime}}{d_{\alpha^\prime}^2 + s}} G_{\alpha^\prime \alpha} |\alpha\> \<\alpha^\prime| }
\end{eqnarray}
And, finally, the expression for $h_\theta$ in the Schmidt basis for region $A$ may be written:
\begin{eqnarray}
h_\theta = h_a - \frac{1}{4} \theta \sum_{\alpha \alpha^\prime}{G_{\alpha^\prime \alpha} \frac{E_{\alpha^\prime} - E_\alpha}{\sinh{[(E_{\alpha^\prime} - E_\alpha)/2]}} |\alpha\> \<\alpha^\prime| } + {\cal{O}}(\theta^2)
\end{eqnarray}
Perturbation theory simplifies further if the operator $O$, written in real space, is broken into pieces that scatter particles inter/intra $A,\barA$:
\be
O \equiv O_{AA} + O_{A \barA} + O_{\barA \barA}
\ee
in obvious notation.   If $L_A << L_\barA$ it is reasonable to keep only the $O_{\barA \barA}$ contribution to $G$ and
\begin{eqnarray}
&h_\theta &= h_a +  \label{pert_h} \\ &-& \frac{1}{4} \theta \sum_{\alpha \alpha^\prime}{  \overline{\<\alpha^\prime|} O_{\barA \barA}+O_{\barA \barA}^\dagger \overline{|\alpha\>} \frac{E_{\alpha^\prime} - E_\alpha}{\sinh{[(E_{\alpha^\prime} - E_\alpha)/2]}} |\alpha\> \<\alpha^\prime| } + {\cal{O}}(\theta^2) \nonumber
\end{eqnarray}
{\bf This is our central result for the modular hamiltonian.} The perturbation to $h_a$ then has the following interpretation: in the $A$ region, the modular hamiltonian matrix element connecting  two Schmidt states $|\alpha\>$ and $|\alpha^\prime\>$ (an "interaction" term in $h_\theta$) is determined by the transition amplitude of $O$ between the corresponding Schmidt states in $\barA$ weighted by a factor involving the "energies" $E^a_\alpha$ of the unperturbed modular hamiltonian. Because of the eigenstate typicality of the pure state $|e_a\>$, the energies  $E^a_\alpha$ {\sl are} approximately energies of an excited state of a noninteracting fermi system restricted to $A$ divided by the effective temperature corresponding to $|e_a\>$.

Recognizing that the Schmidt states $\overline{|\alpha\>}$ are the projections  $\overline{|\alpha\>} = \<\alpha |e_a\>$, the matrix element appearing in (\ref{pert_h}) may be rewritten using
\be
 \overline{\<\alpha^\prime|} O_{\barA \barA}^\dagger \overline{|\alpha\>} = \<e_a|\alpha^\prime\>\<\alpha|e_b\>
\label{G_matrix}
\ee
where it is understood that the r.h.s. is an inner product computed in the $\barA$ hilbert space. The Schmidt states $|\alpha\>$ are eigenvectors of $h_a$ which, according to typicality, is approximately proportional to a free fermi hamiltonian having particles with pseudo-momentum values $p_m \approx 2\pi m/L_A$.  Adopting the notation used for the state $|e_a\>$ itself, the states $|\alpha\>$ are approximately excited states of a free fermi subsystem defined within the $A$ hilbert space:
\be
|\alpha \> = \prod_{m=0}^{M-1} c_{p^\alpha_m}^\dagger |0\>
\label{explicit_alpha}
\ee
 where the set $\{p^\alpha_m \}$ are drawn from the allowed values $p_m$ and the typical number of particles, $M$, in region $A$ should reflect the fixed filling fraction of the excited states $|e_a\>$ and $|e_b\>$. Since both $|e_a\>$ and $|\alpha\>$ can, in principle, be written in real space lattice coordinates, the matrix element (\ref{G_matrix}) can be written in a complicated analytic form, an approach we will follow later. 
 
 \begin{figure}[htbp]
    \centering
    \includegraphics[width=0.6\linewidth]{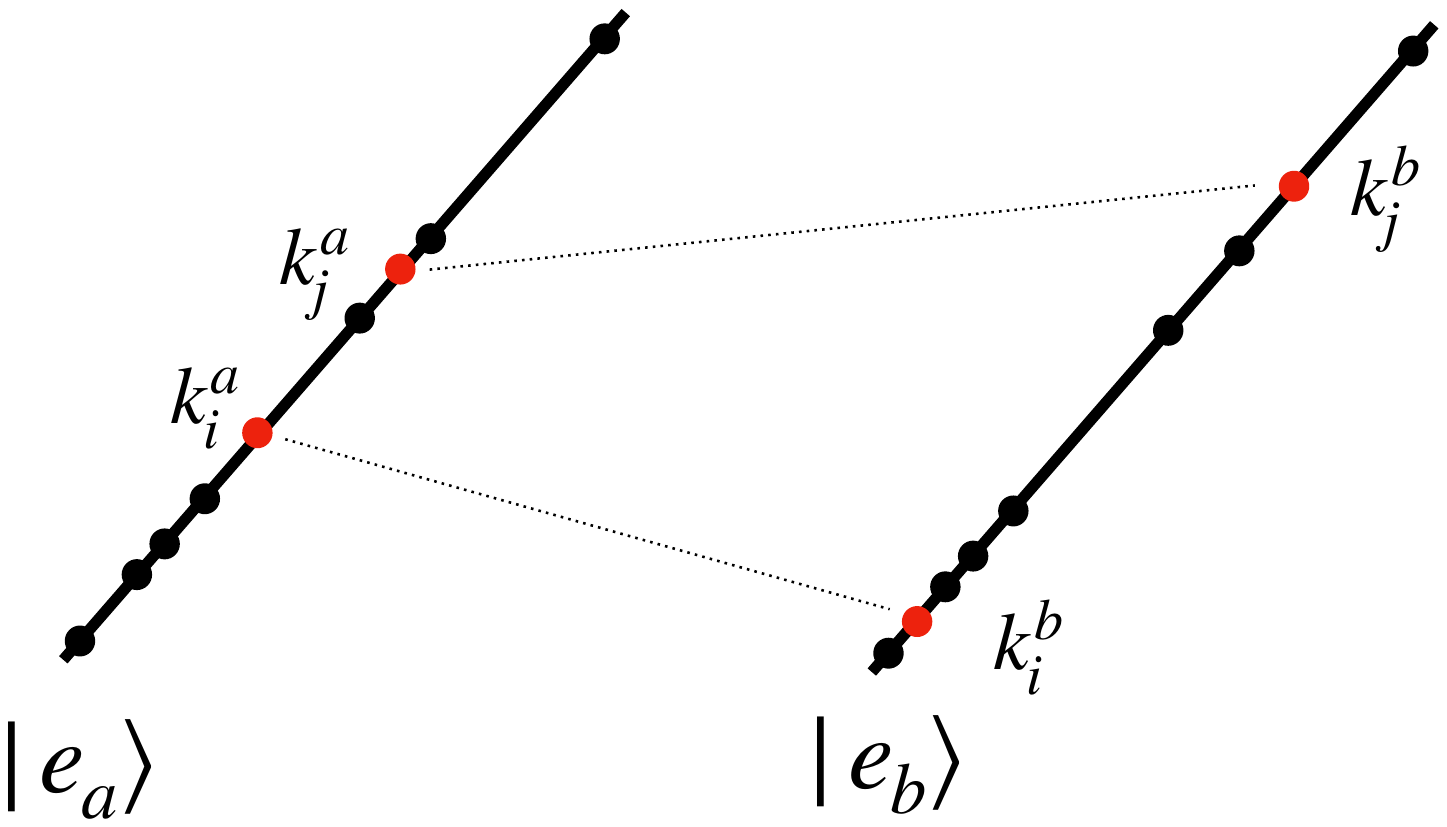}
    \caption{Right moving branches of two degenerate states $|e_a\>$ and $|e_b\> = O^\dagger |e_a\>$ depicting $O^\dagger$ as a two-body operator.}
    \label{momenta}
\end{figure}

Presently, we focus on an estimate of the magnitude of the interacting matrix elements that destroy the gaussian property of $h_\theta$.  First define the part of the excited state that is common to $|e_a\>$ and $|e_b\>$
\begin{eqnarray}
|e\> \equiv \prod_{m=0}^{N-3}{c_{k_m}^\dagger |0\> } = \sum_{{\mathbf{x}} \in A\barA}{D_{{\mathbf{x\cdot k} }} c_{x_0}^\dagger \ldots c_{x_{N-3}}^\dagger |0\> }
\end{eqnarray}
where $D_{{\mathbf{x\cdot k} }} \equiv L^{-N/2+1}\exp{(-i {\mathbf{x\cdot k} })} $ and the $N-2$ elements of $\{k_m \}$ are the common values of the sets $\{k^a_m \}$ and $\{k^b_m \}$. Choosing a specific two-body operator $O^\dagger = c^\dagger_{k^b_i} c^\dagger_{k^b_j} c_{k^a_i} c_{k^a_j}$ and thus the specific set of four momenta that distinguish states $|e_a\>$ and $|e_b\>$,
\be
|e_a\> \equiv c_{k^a_i}^\dagger c_{k^a_j}^\dagger|e\> \,\, {\rm and} \,\, |e_b\> \equiv c_{k^b_i}^\dagger c_{k^b_j}^\dagger|e\> 
\ee
the magnitude of the matrix elements may be approximated using (\ref{G_matrix}). If the $A$ subregion vanishes, the matrix elements vanish because the two degenerate states are orthogonal, $\<e_a|e_b\> =0$.   The zero overlap in this case may be formally expressed
\begin{eqnarray}
\<e_a|e_b\> &=& \frac{1}{L^2} \sum_{x_i^a x_j^a x^b_i x^b_j} e^{ik^a_i x^a_i} e^{ik^a_j x^a_j} e^{-ik^b_i x^b_i} e^{-ik^b_j x^b_j} \times \nonumber \\  &\times& \<e| c_{x^a_i} c_{x^a_j} c^\dagger_{x^b_i} c^\dagger_{x^b_j} |e\> 
\end{eqnarray}
where $c_x \equiv L^{-1/2} \sum_k c_k\exp{(i k x)}$ and $x = 0, \ldots, L-1$.  Note that the excited state $|e\>$ {\sl does} have unique momentum ordering and thus satisfies Wick's theorem. The $\{x_i^a, x_j^a \}$ sum over the entire real space lattice create the expected delta functions of momenta. Introducing relative coordinates for the position and momentum, $\Delta k \equiv k^a_i - k^b_i$ and $\Delta x_{i/j} \equiv x^b_{i/j} - x^a_{i/j} $, we have further assumed that $O^\dagger$ is restricted to either the left-moving or right-moving linearized spectra so that energy degeneracy between $|e_a\>$ and  $|e_b\>$ (in one dimension) implies that $ k^a_j - k^b_j = -\Delta k$ (see figure (\ref{momenta})). Note that the choices for $k^a_i, k^a_j, k^b_i, k^b_j$ are arbitrary, so we will examine just one of the two possible contractions of the four real space operators: $\<e| c^\dagger_{x_i^b} c_{x_i^a}  |e\> \<e| c^\dagger_{x_j^b} c_{x_j^a}  |e\> $. We first examine the $\Delta x_i$ contraction factor contribution to the overlap (the $\Delta x_j$ contraction factor is identical), 
\begin{eqnarray}
& \frac{1}{L}&\sum_{\Delta x_i, x_i^a}{\<e|c^\dagger_{x_i^a + \Delta x_i} c_{x^a_i} |e\> e^{i k^a_i \Delta x_i} e^{i \Delta k x^a_i } } = \nonumber \\
& \frac{1}{L}& \sum_{x_i^a}{e^{i \Delta k x^a_i } } \sum_{\Delta x_i}{g(\Delta x_i) e^{i k^a_i \Delta x_i} }
\label{overlap_factor}
\end{eqnarray}
where $g(\Delta x)$ is the real space fermion correlation function. Both sums in equation (\ref{overlap_factor}) are zero: the first is a delta function $L \delta_{\Delta k, 0}$, which is zero for any two distinct states $|e_a\>$ and  $|e_b\>$; the second sum is the momentum distribution function $n_{k^a_i} = 0$ simply because the momentum state $k^a_i$ is unoccupied in $|e\>$. Thus the overlap $\<e_a|e_b\> =0$ as expected.
 \begin{figure}[htbp]
    \centering
    \includegraphics[width=0.8\linewidth]{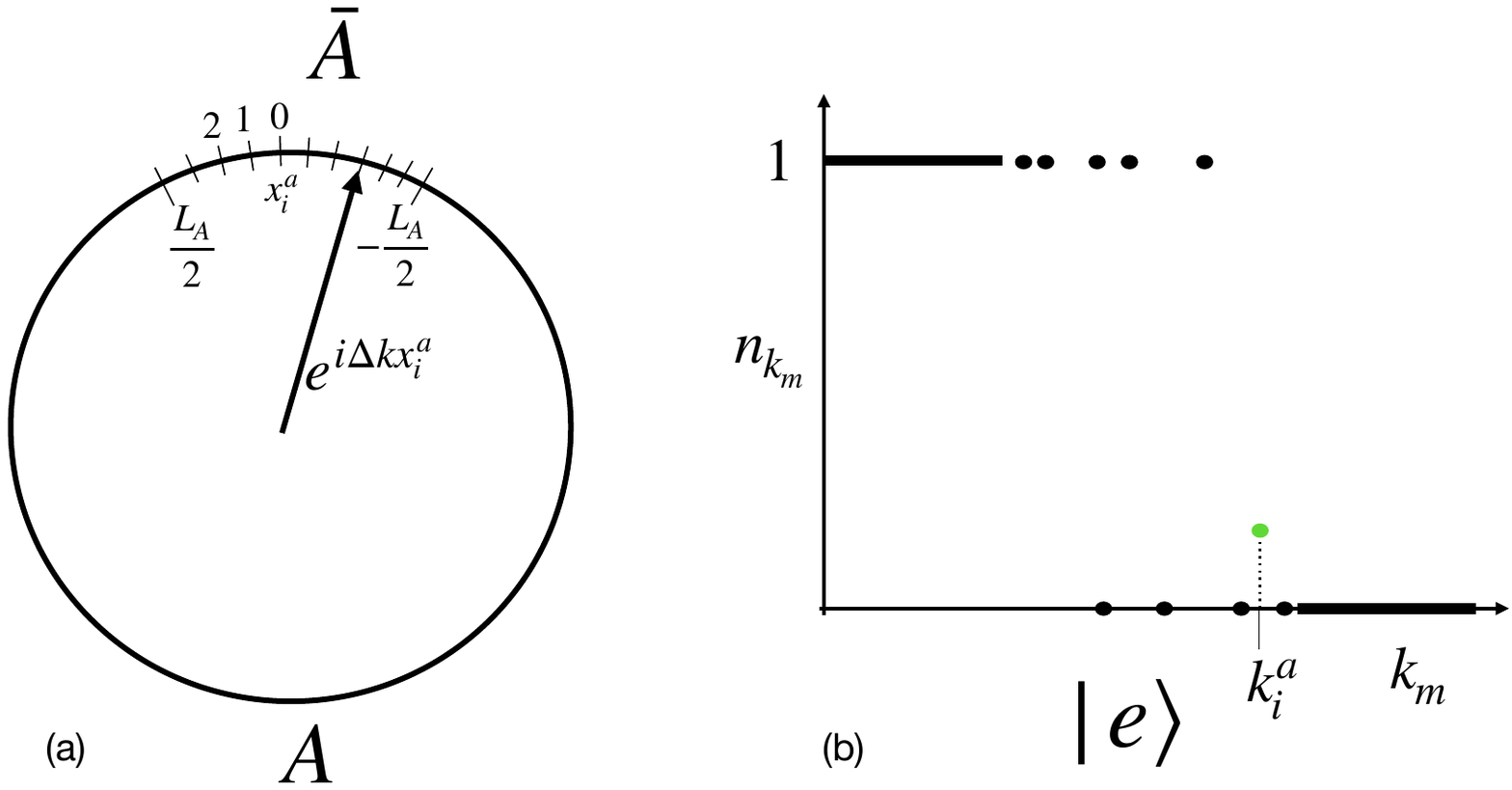}
    \caption{(a) Depiction of Kronecker delta function sum with region $A$ removed following equation (\ref{overlap_factor}). States in the subregion $\barA$ that are antipodal to the excluded subregion $A$ add coherently to the delta function sum. (b) Momentum distribution for a typical state $|e\>$ The state $k^a_i$ is unoccupied in $|e\>$ but, by assumption of typicality of $|e_{a,b}\>$, lies close to occupied states in $|e\>$. Thus $n_{k^a_i}$ becomes nonzero once the states in $A$ are removed from the momentum distribution sum.}
    \label{distribution}
\end{figure}
When the $A$ subregion is expanded to a finite number of sites and those sites become excluded from the sums appearing in the overlap, both factors become non-zero for different reasons. We are justified in choosing $\Delta k << 1/L_A$ as $|e_a\>$ and $|e_b\>$ may be thought of as two neighboring microstates in the sense of the discussion following equation (\ref{corr_func}). In this case, the phase $\Delta k x^a_i$ is nearly constant over $x^a_i \in A$. The contribution to the delta function sum from those sites in the subregion $\barA$ that are antipodal to the excluded subregion $A$ add coherently and generate a factor of ${\cal O}(L_A/L)$ as suggested by figure (\ref{distribution}a). 

Turning to the second sum, we are trying to estimate the nonzero value that $n_{k^a_i}$ achieves when region $A$ is projected out. Recognizing that the single particle momentum distribution function is
\be
n_{k_m} = \sum_{\Delta x}{g(\Delta x)e^{-i\Delta x k_m}}
\ee
the distribution function with region $A$ projected out is may be written:
\be
n^\prime_{k_m} =\frac{1}{L} \sum_{\Delta x \in \barA}{\sum_{\{k_n\}\in |e\>}{n_{k_n}e^{-i\Delta x (k_m-k_n)}}}
\label{mod_distribution}
\ee
Since $\sum_{\Delta x}{\exp{-i\Delta x(k_m-k_n)}} = L\delta_{m,n}$ this expression is identically satisfied for the $\Delta x$ sum running over all $A\barA$. In the particular state $|e\>$, the momentum state $k^a_i$ is unoccupied, but typically lies close to occupied states in that the set of typical states---including $|e_{a}\>$ and $|e_{b}\>$---are generated by permuting momentum states in the transition region where typical occupancies shift from $1$ to $0$.  We argue that it is then sufficient to estimate the modification to the momentum delta function suggested in equation (\ref{mod_distribution}) for the particular momentum state $k^a_i$ by assuming that it is {\sl adjacent} to an occupied state $k^a_{i+1}$. Setting $k_m = k_i^a$ in equation (\ref{mod_distribution}) and using  $k^a_{i+1} - k^a_i = 2\pi/L$,  
\be
n_{k^a_i} \approx \frac{1}{L} |  \sum_{\Delta x \in \barA}{e^{-2i\pi \Delta x/L}}   | \sim {\cal O}(\frac{L_A}{L})
\ee
where we have applied the argument used for the first sum in equation (\ref{overlap_factor}) to approximate the present sum. 
We conclude that the "projected" contractions that appear in $\<e_a|\alpha^\prime\>\<\alpha|e_b\>$ may be approximated by
\be
\frac{1}{L}\sum_{\Delta x_i, x_i^a \in \barA}{\<e|c^\dagger_{x_i^a + \Delta x_i} c_{x^a_i} |e\> e^{i k^a_i \Delta x_i} e^{i \Delta k x^a_i } } \sim {\cal O}(\frac{L_A^2}{L^2})
\label{contraction_estimate}
\ee
Since $\<e_a|\alpha^\prime\>\<\alpha|e_b\>$ involves the product of two contractions, our final estimate for the magnitude of two-body (interacting) matrix elements of the modular hamiltonian (\ref{pert_h}) is 
\be
\overline{\<\alpha^\prime|} O_{\barA \barA}^\dagger \overline{|\alpha\>} =  \<e_a|\alpha^\prime\>\<\alpha|e_b\> \sim {\cal O}(\frac{L_A^4}{L^4})
\label{h_estimate}
\ee

From (\ref{G_matrix}), the matrix elements of the modular hamiltonian may also be explicitly calculated, but the expression is not particularly enlightening. Through eigenstate typicality, the states $|\alpha\>$ are thought to resemble typical excited states constructed in subregion $A$ and expressed in momentum space following (\ref{explicit_alpha}). To express the desired matrix elements they must be transformed to real space; to that end, the real space creation and annihilation operators within region $A$ are related to the pseudo-momentum ones in the usual fashion: 
\be
c_{p_m} = L_A^{-1/2} \sum_{y=0}^{L_A-1}{ c_y e^{-i p_{m} y } } =: \sum_{y=0}^{L_A-1}{ c_x \phi_{ym} }
\ee
For $M$ particles in region $A$, it follows that $|\alpha\>$ representing a particular excited state described by the set of occupied pseudo-momenta $\{p_m \}$ is written:
\be
|\alpha\> = \sum_{\mathbf{y}}{A_{\mathbf{y}} \prod_{m=0}^{M-1}{ c_{y_m}^\dagger} |0\>}
\ee
where each $y_m$ (running over sites in  $[0,L_A-1] \in A$)  of the vector $\mathbf{y}$ is a spatial coordinate paired with the pseudo-momentum $p_m$ and $A_{\mathbf{y}}$ is a matrix formed from the products of $\phi_{ym}$. As before, the excited states $|e_{a/b}\>$ may be expressed as real space operators acting upon $|e\>$:
\be
|e_{a/b}\> = \sum_{x^{a/b}_{1/2} \in A\barA}{T^{a/b}_{x_1^{a/b} x_2^{a/b}}c^\dagger_{x_1^{a/b}} c^\dagger_{x_1^{a/b}} }|e\>
\ee
where the $T$ matrices are formed from products of free fermion wave-functions.
Finally the expression (\ref{G_matrix})  which enters the modular hamiltonian matrix element (\ref{pert_h}) may be written
\begin{eqnarray}
 &&\overline{\<\alpha^\prime|} O_{\barA \barA}^\dagger \overline{|\alpha\>} = \<e_a|\alpha^\prime\>\<\alpha|e_b\> = \nonumber \\
 &&\sum_{x^{a/b}_{1/2} \in A\barA} T^{a}_{x_1^{a} x_2^{a}} T^{b}_{x_1^{b} x_2^{b}} \sum{A_{\mathbf{y}^{\alpha}}^{\alpha} A_{\mathbf{y}^{\alpha^\prime}}^{\alpha^\prime} } \times \\ && \<e|c_{x_1^a} c_{x_2^a} \prod_{m=0}^{M-1}{c_{y^{\alpha}_m}} \prod_{m=0}^{M-1}{  c^\dagger_{y^{\alpha^\prime}_m}}   c^\dagger_{x_1^b} c^\dagger_{x_2^b}  |e\>
\end{eqnarray}
Since expectation values within the pure, excited state $|e\>$ satisfy Wick's theorem, this expression involves products of contractions similar to those of the modular hamiltonian magnitude estimate (\ref{contraction_estimate}). Thus the perturbative corrections to the modular hamiltonian corresponding to the weak mixture of two pure states is in principle computable, but the expression is complicated.

\section{Conclusion and Outlook}

We believe this calculation has intrinsic interest for two reasons: (1) it demonstrates that the reduced density matrix of typical states of free fermions will exhibit deviations from the fermi-dirac distribution, representative of effective interactions brought about by the defining wave function being a mixture of pure states.  Such states violate Wick's theorem and thus leave open the possibility of a non-gaussian modular hamiltonian.  (2) It is a the direct construction of a modular hamiltonian which demonstrates the anticipated---but largely unexplored---features of non-locality and interactions.

We now turn to the question of the physical relevance of this result. The weak, repulsive interactions found in realistic fermi systems are irrelevant under RG flow and thus fermions are typically asymptotically free at low energies. Through the argument of Lai and Yang, one would expect the modular hamiltonian of a small subsystem to be approximately noninteracting and local.  In terms of the ETH, the renormalized weak interactions in the RG approach provide the necessary quantum chaos to equilibrate realistic fermions, in contrast with the integrable case of true noninteracting fermions which relaxes to the generalized Gibbs ensemble (see, for instance \cite{2016JSMTE..06.4002E}). Here we briefly review the ETH argument and discuss how our present result and the results of Lai and Yang might be meaningfully unified.   

Consider a set of weakly repulsive identical fermions on a lattice described by a hamiltonian $H$ and its complete set of energy eigenstates $\{ |\ \psi_a\>\}$. The fermions are prepared in a nonequilibrium (e.g. spatially inhomogeneous) initial many body state with energy $E_0$. This state may be written
\be
|\Psi(0)\> = \sum_a c_a  |\ \psi_a\>
\label{init}
\ee
where $\{ c_a \}$ carry the information of the initial condition. The state evolves as  
\be
|\Psi(t)\> = \sum_\alpha c_a e^{-i E_a t} |\ \psi_a\>
\ee
and it is seen that as $t \rightarrow \infty$ the expectation value of a local operator $\hat{P}$ serving as a probe for equilibrium realizes the diagonal distribution if there are no perfect degeneracies:
\be
\label{diag_dist}
\< \hat{P} \> = \sum_a |c_a|^2  \<\psi_a | \hat{P} |\psi_a\>
\ee
The dominant coefficients $c_a$ of the sum will correspond to total system energies $E_a \approx E_0$, and the ETH follows from the observation that $\< \psi_a | \hat{P} | \psi_a\>$ is largely independent of the index $a$ for the dominant terms. Thus, typical excited eigenstates within a small range of energy all yield very similar expectation values and the matrix element in equation (\ref{diag_dist}) comes out of the sum.  Since the state is normalized, $\sum |c_a|^2 \approx 1$ and the expectation values $\<\hat{P}\>$ reflects any typical state $|\psi_{\rm T}\>$ from the set  $\{ |\ \psi_a\>\}$,
\be
\<\hat{P}\> = \<\psi_{\rm T}| \hat{P} |\psi_{\rm T}\>
\label{ETH}
\ee
 and is independent of the particular choice of initial conditions expressed by $\{ c_a\}$. 

It is reasonable to assume that in a such a realistic Fermi system, the energy eigenstates that contribute to the diagonal distribution would be complicated superpositions of the underlying quasi-degenerate, asymptotically free fermi states. In this scenario, an initial condition of specified energy $E_0$ would have a dominant projection upon a large set of asymptotically free, quasi-degenerate states representing the fermi surface. To this end we model the low energy states $|\psi_a\>$ contributing to the diagonal distribution in a fermi system as having large---possibly extensive---degeneracies. Consider the $\{|\psi_a\>\}$ that dominantly contribute to the diagonal distribution to be perfectly degenerate (rather than quasi-degenerate centered about $E_0$ as is found computationally \cite{2008Natur.452..854R}). To compute the expectation value of a local operator such as $\hat{P}$, it is sufficient to compute the RDM for a subsystem $A$ of the state (\ref{init})
\begin{eqnarray}
\rho &=& \tr_{\barA} \sum_{ab}^N{c_a c_b^* |\psi_a\>\<\psi_b|}  \\
&=& \tr_\barA \sum_a^N{|c_a|^2 |\psi_a\>\<\psi_a|} + \tr_{\barA} \sum_{a\ne b}^N{c_a c_b^* |\psi_a\>\<\psi_b|} \\
&\approx& \rho_{\rm T} + \frac{1}{N} \tr_\barA \sum_{a\ne b}^N{\epsilon_{ab} |\psi_a\>\<\psi_b|} 
\label{ETH_degenerate}
\end{eqnarray}
In the last line, we have assumed the typicality of the $N$ degenerate states $\{|\psi_a\>\}$ and based on the expected normalization, $c_a \sim N^{-1/2}$, rewritten the coefficients as an ${\cal O}(1)$ matrix: $\epsilon_{ab}$.  Now the first term of (\ref{ETH_degenerate}) may be thought of as realizing the ETH in the standard way of equation (\ref{ETH}):
\be
\<\hat{P}\> = \<\psi_{\rm T}| \hat{P} |\psi_{\rm T}\> = \tr_A(\rho_{\rm T}\hat{P})
\ee
where $\rho_{\rm T}$ is a typical reduced density matrix computed from a selected member of the degenerate set of states.

The second term of (\ref{ETH_degenerate}) may be treated as a perturbation along the lines of calculation in section IV. If the first term reflects the free fermi calculation of $\< \hat{P} \>$ at a temperature dependent upon $ E_0$, the second term would yield a correction to $\< \hat{P} \>$ based upon interacting corrections to the modular hamiltonian following the treatment of equation (\ref{h_theta_def}). The modular hamiltonian $h_\theta \equiv -\log{\rho_\theta}$ in (\ref{h_theta_def}) was found to have additive corrections of ${\cal O}(\theta)$ given by (\ref{pert_h}). Therefore, the sum appearing in (\ref{ETH_degenerate}) also gives rise to additive corrections to $h \equiv -\log{\rho}$:
\begin{eqnarray}
h &=& h_{\rm T} + \\
&-& \frac{1}{N} \int_0^\infty{[(\rho_{\rm T}+ sI)^{-1}  [\tr_\barA \sum_{a\ne b}^N{\epsilon_{ab} |\psi_a\>\<\psi_b|}] (\rho_{\rm T} + sI)^{-1}) ]ds}
\label{corrections_mutli_degeneracy}
\end{eqnarray}
To reach an approximate expression analogous to (\ref{pert_h}) we first define the Schmidt basis for each member of $\{|\psi_a\>\}$ and the operator that connects different members of the degenerate set:
\begin{eqnarray}
|\psi_a\> &=& \sum{d^a_\alpha |\alpha^a\> |\bar{\alpha^a}\>} \\
|\psi_b\> &=& O^\dagger_{ab} |\psi_a\>
\end{eqnarray}
Now we widely apply "typicality" and assume that $\rho_{\rm T}$ approximately shares common eigenvalues with all of the states in $\{|\psi_a\>\}$: $\rho_{\rm T} |\alpha^a\> \approx d_\alpha^2 |\alpha^a\>$. This is not an unreasonable assumption in that one expects $d_\alpha^2 = \exp{(-E_\alpha)}$ to reflect the energies of the fermi sea subsystem independent of the typical degenerate state chosen.   

With these assumptions, the estimate of the magnitude of the two-body modular hamiltonian matrix elements proceeds as before.  Based upon (\ref{h_estimate}) the $N^2$ terms in the sum (\ref{ETH_degenerate}) would yield a perturbative correction of order ${\cal O} (N (L_A/L)^\delta) $ where $\delta$ depends upon the spatial dimension and the detailed form of the many-body operators---the counterparts of $O$---connecting the states $\{|\psi_a\>\}$ (we have found $\delta = 4$ for $d=1$ and two-body states). We argue that these details do not alter the conclusion that the magnitude of the of the modular hamiltonian interacting matrix elements are {\sl algebraic} in $L_A/L$. The ultimate magnitude of these corrections then depends upon the number of degenerate states contributing, with $N \sim L/L_A$ saturating perturbation theory.

Note that the corrections to the modular hamiltonian would depend implicitly upon the set $\{c_a \}$ expressing the initial conditions and thus the information of the initial conditions would be encoded in correlations between nominally free fermions measured within $A$. Within this picture it is important to distinguish between two different potential sources of deviation from pure fermi statistics.  First, there are the familiar perturbative corrections to asymptotically free fermions that would be brought about by residual interactions---those which presumably thermalize a realistic weakly interacting fermi system.  In a standard QFT treatment these are commonly termed quasiparticle lifetime effects following from inelastic scattering. In this manuscript, we conjecture that there is a second source of deviations from pure fermi statistics---interacting corrections to the modular hamiltonian---expressing unmeasured information of the initial conditions. This information could be partially reconstructed, in principle, by measuring multiparticle fermion correlation functions in a subregion of the bulk. This scheme is, of course, wholly impractical experimentally; rather our conclusion concerns the final fate of information about initial conditions.  



We believe that the detailed form of the modular hamiltonian deserves further study. With a large, arbitrary superposition of degenerate states and $L_A$ sufficiently small, the modular hamiltonian likely takes the form of a random interacting matrix resembling that of the Sachdev-Ye-Kitaev model \cite{2022RvMP...94c5004C}. However, it is important to recognize that the well-studied dynamics of such models appearing in the study strange metals and non-fermi liquids---applied to the the present case---would correspond to modular {\sl flow} of the modular hamiltonian rather than the real time evolution resulting from the bulk hamiltonian. 

Lastly, we comment on a possible extension of our work to the Strong ETH or Subsystem ETH \cite{2018PhRvE..97a2140D} which postulates that the modular hamiltonian, $h$, of a subsystem is approximately proportional to the bulk hamiltonian, $H$. The low energy spectrum of an interacting bulk hamiltonian, $H$, of a fermi system might be characterized by degenerate perturbation theory and the modular hamiltonian, $h$, constructed from a low energy mixture of degenerate states. In degenerate perturbation theory, the effective $O$ operators coupling degenerate states appearing in our calculation are simply computed from $H$ and the free particle energy spectrum. If an argument could be found for how scattering from $O$ is projected onto the real space scattering in subsystem $A$, the perturbative arguments presented here might be used to construct $h$ and demonstrate its equivalence to the bulk hamiltonian $H$ \cite{2020arXiv201205248P}. 

\appendix

\section{Logarithm Matrix Expansion and the Fr{\'{e}}chet Derivative}

Perturbation theory for the modular hamiltonian involves expanding the logarithm of a matrix perturbed by another small matrix (equation (\ref{log_expansion})). The expansion to ${\cal{O}}(\epsilon)$ may be found directly by matrix manipulations of the integral expression for the logarithm of a matrix resulting in (\ref{log_expansion}).  Alternatively, (\ref{log_expansion}) may be derived by using a Taylor expansion based upon the Fr{\'{e}}chet derivative, a type of directional derivative for a matrix valued function.

Let $F[X]$ be a matrix valued function of a matrix $X$. The Taylor expansion of $F[X+\epsilon Y]$ may be formally written:
\be
F[X+\epsilon Y] = F[X] +  D_Y F[X] \cdot \epsilon Y + {\cal{O}}(\epsilon^2)
\ee
where $D_Y$ is the Fr{\'{e}}chet derivative representing the derivative in the "direction" of matrix $Y$ and $\epsilon$ is a small parameter. Further we use 
\be
\delta_Y F = D_Y F[X] \cdot \epsilon Y
\ee 
to denote the variation of the function $F[X]$ in the $Y$ - direction. Since the integral representation of the matrix logarithm involves
\be
\log{X} = A -\lim_{ \Lambda \rightarrow \infty}  \int_0^\Lambda{(X + s I)^{-1}}
\ee
where $A$ is a constant matrix,
\begin{eqnarray}
\log{[X+\epsilon Y]} - \log{[X]} &=& D_Y[\log{X}] \vert_X \cdot \epsilon Y +\dots \\
&=& -  \lim_{ \Lambda \rightarrow \infty}  \int_0^\Lambda{ D_Y(X + s I)^{-1} \cdot \epsilon Y} + \ldots
\end{eqnarray}
Rather than $\delta_Y (X + s I)^{-1}$, consider the evaluation of the variation of $1/X$, $\delta_Y X^{-1}$. Noting that
\be
X^{-1} X = I \Longrightarrow \delta_Y(X^{-1} X) = 0
\ee
Then,
\be
\delta_Y X^{-1} = -X^{-1} (\epsilon Y) X^{-1}
\ee
and
\be
 \lim_{ \Lambda \rightarrow \infty}  \int_0^\Lambda{ D_Y(X + s I)^{-1} \cdot \epsilon Y} = -\epsilon  \lim_{ \Lambda \rightarrow \infty}  \int_0^\Lambda{ (X + s I)^{-1}   Y (X + s I)^{-1}  }
\label{frechet}
\ee
Equation (\ref{log_expansion}) follows directly from (\ref{frechet}).

\end{document}